\documentclass[a4paper,fleqn,usenatbib]{mnras}

\usepackage{newtxtext,newtxmath}

\usepackage[T1]{fontenc}
\usepackage{ae,aecompl}

\usepackage{graphicx}	% Including figure files
\usepackage{amsmath}	% Advanced maths commands
\usepackage{amssymb}	% Extra maths symbols

\defcitealias{Hach99}{H99}

\title[He-enriched MS donor scenario for SNe Ia]{Can the Helium-Enriched Main-Sequence Donor Scenario Hide Enough Hydrogen to Explain Type I\lowercase{a} Supernovae?}

\author[Z.-W. Liu and R.~J. Stancliffe]{
Zheng-Wei Liu${}$\thanks{E-mail: liuzheng-wei@hotmail.com}
and Richard J. Stancliffe${}$
\\
Argelander-Institut f\"ur Astronomie, Auf dem H\"ugel 71, D-53121, Bonn, Germany\\
}

\date{Accepted 2017 May 17. Received 2017 May 16; in original form 2017 April 24}

\pubyear{2017}

\begin{document}
\label{firstpage}
\pagerange{\pageref{firstpage}--\pageref{lastpage}}
\maketitle

% Abstract of the paper
\begin{abstract}
Hydrodynamical simulations predict that a large amount of hydrogen  ($\gtrsim0.1\,M_{\sun}$) is removed from a hydrogen-rich companion star by the SN explosion in the single-degenerate scenario of Type Ia supernovae (SNe Ia). However, non-detection of hydrogen-rich material in the late-time spectra of SNe Ia suggests that the hydrogen mass stripped from the progenitor system is $\lesssim0.001$--$0.058\,M_{\sun}$. In this letter we include thermohaline mixing into self-consistent binary evolution calculations for the helium-enriched main-sequence (HEMS) donor channel of SNe Ia for the first time. We find that the swept-up hydrogen masses expected in this channel are around 0.10-0.17$\,M_{\sun}$, which is  higher than the observational limits, although the companion star is strongly helium-enriched when the SN explodes. This presents a serious challenge to the HEMS donor channel.

\end{abstract}


%%%%%%%%%%%%%%%%%%%%%%%%%%%%%%%%%%%%%%%%%%%%%%%%%%

%%%%%%%%%%%%%%%%% BODY OF PAPER %%%%%%%%%%%%%%%%%%

\section{Introduction} \label{sec:intro}

Type Ia supernovae (SNe Ia) play a fundamental role in astrophysics.  However, despite recent progress on both the theoretical and observational side, their specific progenitor systems have not yet been identified \citep[][]{Hillebrandt2000, Maoz13}. The two favored classes of SN Ia progenitors are the single-degenerate (SD) and double-degenerate (DD) scenario. In the DD scenario, two carbon-oxygen white dwarfs (C+O WDs) merge due to gravitational wave radiation, triggering an SN Ia explosion \citep{Iben84, Webb84}. In the SD scenario, the WD accretes matter from either a main-sequence (MS), subgiant (SG), red-giant (RG), or helium star companion, to ignite a thermonuclear explosion when its mass reaches the Chandrasekhar-mass limit \citep{Whel73, Han04}.

In the SD scenario, ejecta from the SN explosion hits the companion star, removing some its outer layers which are either hydrogen- or helium-rich depending on the nature of the companion \citep{Whee75}. If a significant amount of hydrogen mass can be blown off from the companion star, some features of hydrogen emission could be detectable in their late-time spectra, depending on the distances to the  observed SNe Ia. To date, no strong evidence of the swept-up hydrogen has been detected.\footnote{Although there is a peculiar class of SNe Ia displaying strong features of hydrogen in their spectra, e.g., SN~2002ic \citep{Hamuy2003, Wood-Vasey2004}, SN~2005gj \citep{Aldering2006} and PTF11kx \citep{Dilday2012}, they have been generally suggested to come from interaction of the SN ejecta with their hydrogen-rich circumstellar medium \citep{Silverman2013}.} It has been claimed that there is an upper limit on the hydrogen-mass in the progenitor systems of SNe Ia, namely $0.001-0.058\,M_{\odot}$  \citep[e.g.,][]{Leon07, Lund13, Lund15, Shap13, shappee16, Graham2015, Maguire2016}. However, two- and three-dimensional hydrodynamical simulations have predicted that about $5\%$--$30\%$ of the companion mass, i.e., $\gtrsim0.1\,M_{\odot}$, can be removed from outer layers of a MS, SG or RG companion star by the SN explosion (e.g., \citealt{Mari00, Pakm08, Liu12, Liu13a, Liu13b, Pan12, Boehner2017}), which is in conflict with the observational constraints on the swept-up hydrogen masses. This is a serious challenge for the SD scenario.

\begin{figure*}
  \begin{center}
    {\includegraphics[width=1.85\columnwidth, angle=360]{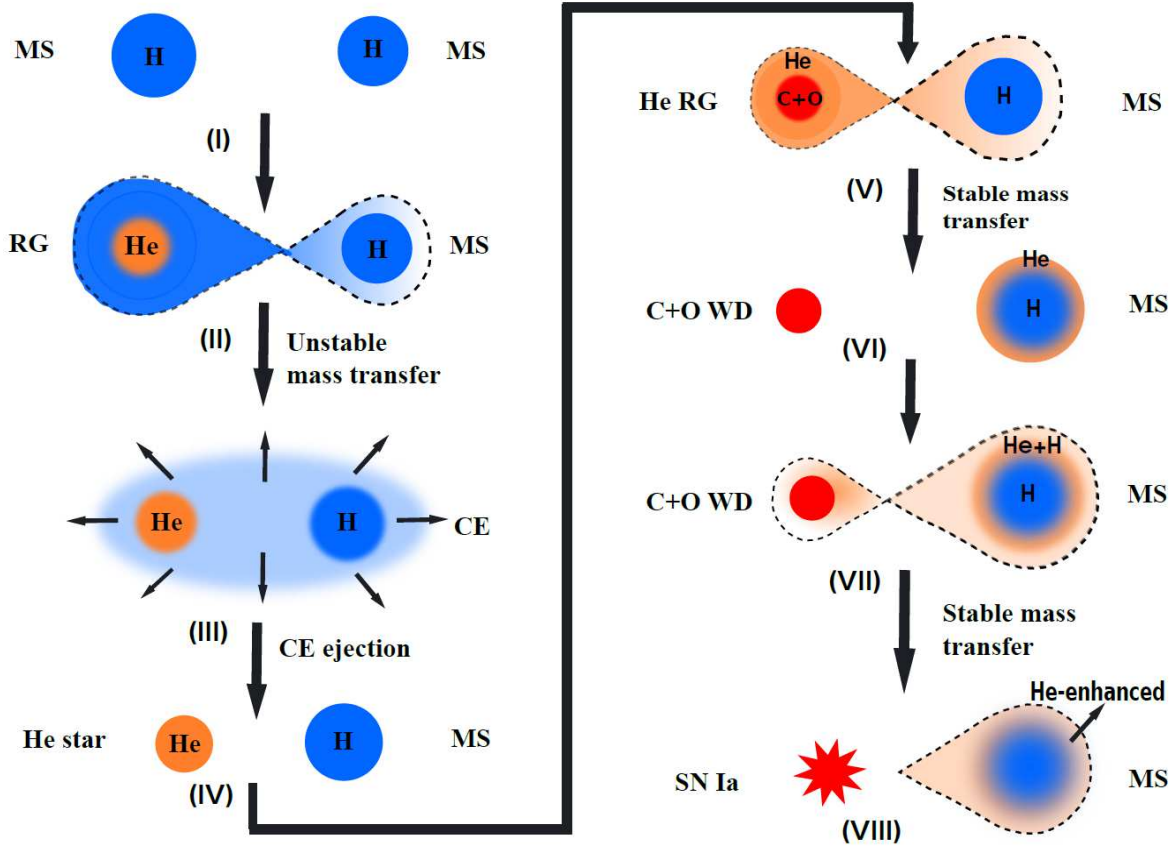}}
    \caption{Binary evolutionary path of the helium-enriched main-sequence donor scenario for SNe Ia. We refer the reader to \citetalias{Hach99} for a detailed description of the system's evolution.}
\label{Fig:1}
  \end{center}
\end{figure*}

In \citet[][hereafter \citetalias{Hach99}]{Hach99}, the helium-enriched MS (HEMS) donor channel was proposed as a new evolutionary path to SNe Ia. In this channel, the binary system undergoes a common envelope event to form a helium star and a MS companion. This helium star subsequently overfills its Roche lobe during core helium burning to deposit helium-rich material onto its MS companion. As a consequence, a binary system consisting of a C+O WD and a helium-enriched MS companion star is produced. This helium-enriched MS star then experiences Roche-lobe overflow (RLOF), transferring matter to the WD, which grows in mass until the Chandrasekhar limit is reached and an SN Ia occurs  (Fig.~\ref{Fig:1}). In this scenario, the MS companion may have strongly helium-enriched outer layers when the SN Ia explodes if a sufficient amount of helium-rich material has been accumulated onto it (and these have not yet been stripped by its own RLOF episode). The material subsequently removed from the companion by the SN explosion could then be hydrogen-deficient, which may provide a way to explain the lack of hydrogen-features in the late-time spectra of SNe Ia. 

Rather than leave a layer of almost pure helium on the surface of the MS companion, the material accreted from the helium star will become mixed into the companion's interior by the action of thermohaline mixing. This process occurs when the mean molecular weight of the stellar gas increases towards the surface, in this case because the accreted helium layer has a higher mean molecular weight than the material of the MS companion. A gas element, displaced downwards and compressed, will be hotter than its surroundings. It will therefore lose heat, increase in density and continue to sink. This results in mixing on a thermal timescale until the molecular weight difference has disappeared \citep{Kippenhahn1980, Stancliffe2007}.  The effects of thermohaline mixing on the structure and composition of stars have been widely analyzed either in low-mass binaries or in massive systems (e.g., \citealt{Wellstein2001, Stancliffe2007, Stancliffe2008, Stancliffe2010, Siess2009}), as thermohaline mixing is expect to naturally occur in binary systems when material that has undergone nuclear processing from the primary star is transferred to its less evolved secondary. At stage $\mathrm{V}$ of the HEMS donor scenario (Fig.~\ref{Fig:1}), helium-rich material from the primary star is transferred to the surface of the MS secondary. This alters the surface composition of the secondary because thermohaline mixing is expected to naturally occur as this accreted material has a greater mean molecular weight than that of the MS secondary \citep{Kippenhahn1980}. Therefore, thermohaline mixing cannot be neglected in the HEMS donor scenario.

In this letter, we perform self-consistent binary evolution calculations for the HEMS donor scenario, including the effects of thermohaline mixing. We use the outcome of these simulations to calculate the  swept-up hydrogen mass expected when the SN ejecta interacts with the MS companion in these systems.

\begin{table*}\renewcommand{\arraystretch}{1.03}
\centering
\caption{Models in our binary evolution calculations with the {\sc STARS} code. We list only some of those systems that produce an SN Ia.}
 \begin{tabular}{@{}lccccccccccc@{}}
 \hline\hline
Model & $M_{\mathrm{MS}}$(IV)  & $M_{\mathrm{He}}$(IV)  &  $P_{\mathrm{orb}}$(IV)  &  $P_{\mathrm{orb}}$(VI) &  $M_{\mathrm{WD}}$(VI) & $M_{\mathrm{2}}$(VIII) & Q(VIII)  & $R_{\mathrm{2}}$(VIII) & $\Delta M_{\mathrm{He}}$(V)  &\ $\Delta M_{\mathrm{H,1}}$ &\ $\Delta M_{\mathrm{H,2}}$ \\
      &  $[M_{\sun}]$            & $[M_{\sun}]$             & $[\mathrm{days}]$             & $[\mathrm{days}]$ & $[M_{\sun}]$                & $[M_{\sun}]$              &                                   
          &  $[R_{\sun}]$             & $[M_{\sun}]$        & \ \ \  $[M_{\sun}]$ & \ \ \  $[M_{\sun}]$ \\
\hline
M01     & 2.0  &  1.0	&   1.0		&  1.3	&  0.84 &  1.40	&  2.64	&  2.2	&	0.16	&\ \ \	0.10	&\ \ \	0.13	\\
M02     & 2.0  &  1.2	&   2.0		&  3.2	&  0.89 &  1.54	&  2.58	&  4.3	&	0.31	&\ \ \ 	0.11	&\ \ \	0.13	\\
M03     & 2.5  &  1.2	&   1.0		&  1.7	&  0.89 &  2.18	&  2.39	&  2.5	&	0.31	&\ \ \ 	0.14	&\ \ \	0.18	\\
M04     & 3.0  &  1.2	&   2.0		&  3.6	&  0.89 &  2.75	&  2.27	&  4.1	&	0.31	&\ \ \	0.16	&\ \ \	0.22	\\
M05     & 2.0  &  1.4	&   2.0		&  3.5	&  0.94 &  1.61	&  2.56	&  5.3	&	0.46	&\ \ \	0.10	&\ \ \	0.13	\\
M06     & 2.5  &  1.4	&   4.0		&  7.8	&  0.95	&  2.42	&  2.34	&  7.5	&	0.45	&\ \ \	0.14	&\ \ \	0.18	\\
M07     & 2.0  &  1.6	&   1.0		&  1.9	&  1.00	&  2.09	&  2.41	&  3.1	&	0.60	&\ \ \	0.12	&\ \ \	0.15	\\
M08     & 2.0  &  1.6	&   3.0		&  5.5	&  1.01	&  2.06	&  2.42	&  6.5	&	0.59	&\ \ \	0.11	&\ \ \	0.15	\\
M09     & 2.5  &  1.6	&   5.0		&  10.5 &  1.01	&  2.65	&  2.30	&  9.9	&	0.59	&\ \ \	0.13	&\ \ \	0.18	\\
M10     & 2.0  &  1.8	&   2.0		&  3.6	&  1.09	&  2.31	&  2.36	&  5.3	&	0.69	&\ \ \	0.12	&\ \ \	0.16	\\
M11     & 2.5  &  1.8	&   4.0		&  8.7	&  1.08	&  2.87	&  2.26	&  9.5	&	0.73	&\ \ \	0.14	&\ \ \	0.19	\\
M12     & 3.0  &  1.8	&   6.0		&  14.5 &  1.08	&  3.38	&  2.18	&  13.9	&	0.72	&\ \ \	0.17	&\ \ \	0.22	\\
\hline       
\end{tabular}

\medskip
\flushleft
\textbf{Note.} $M_{\mathrm{MS}}$\,(IV), $M_{\mathrm{He}}$\,(IV) and  $P_{\mathrm{orb}}$\,(IV) are the mass of the MS star, mass of the helium star and orbital period of the binary system at stage IV of Fig.~\ref{Fig:1}. $P_{\mathrm{orb}}$\,(VI) and $M_{\mathrm{WD}}$(VI) are the orbital period of the system and the mass of the WD at stage VI. $M_{\mathrm{2}}$\,(VIII) and $R_{\mathrm{2}}$\,(VIII) are the mass and radius of the companion star at stage VIII, and Q\,$=A_{\mathrm{orb}}/R_{\mathrm{2}}$ is the ratio of orbital separation to companion radius. Here, $\Delta M_{\mathrm{He}}$\,(V) is the total helium masses that are transferred to a main-sequence companion star at stage V. The amount of swept-up hydrogen masses owing to the SN explosion, $\Delta M_{\mathrm{H,1}}$ and $\Delta M_{\mathrm{H,2}}$, are estimated based on recent hydrodynamical simulations of \citet{Liu12} and \citet{Pan12}, respectively. \\
\label{Table:1}
\end{table*}

\begin{figure*}
  \begin{center}
    {\includegraphics[width=0.9\columnwidth, angle=360]{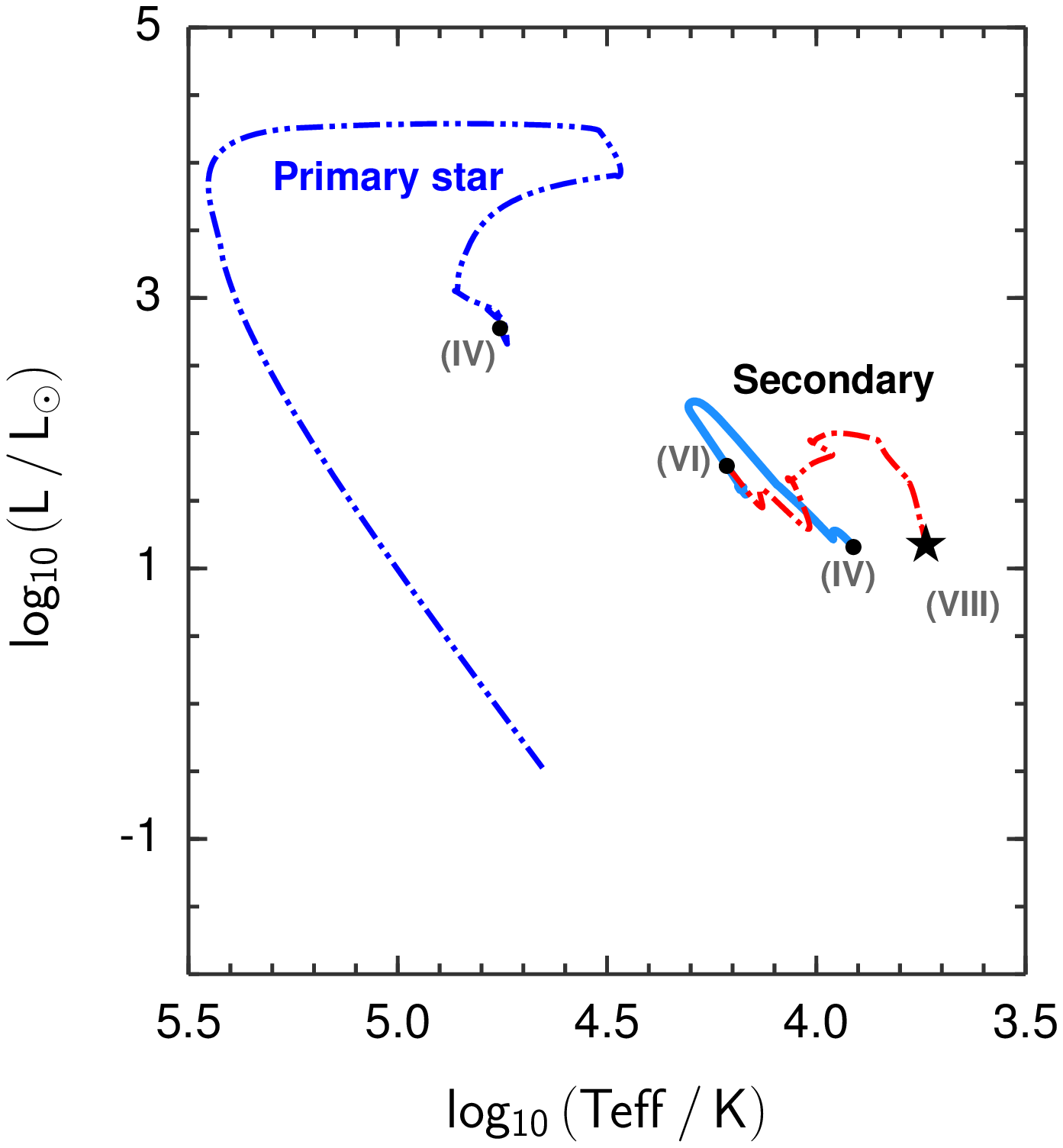}}
    \hspace{0.2in}
    {\includegraphics[width=0.9\columnwidth, angle=360]{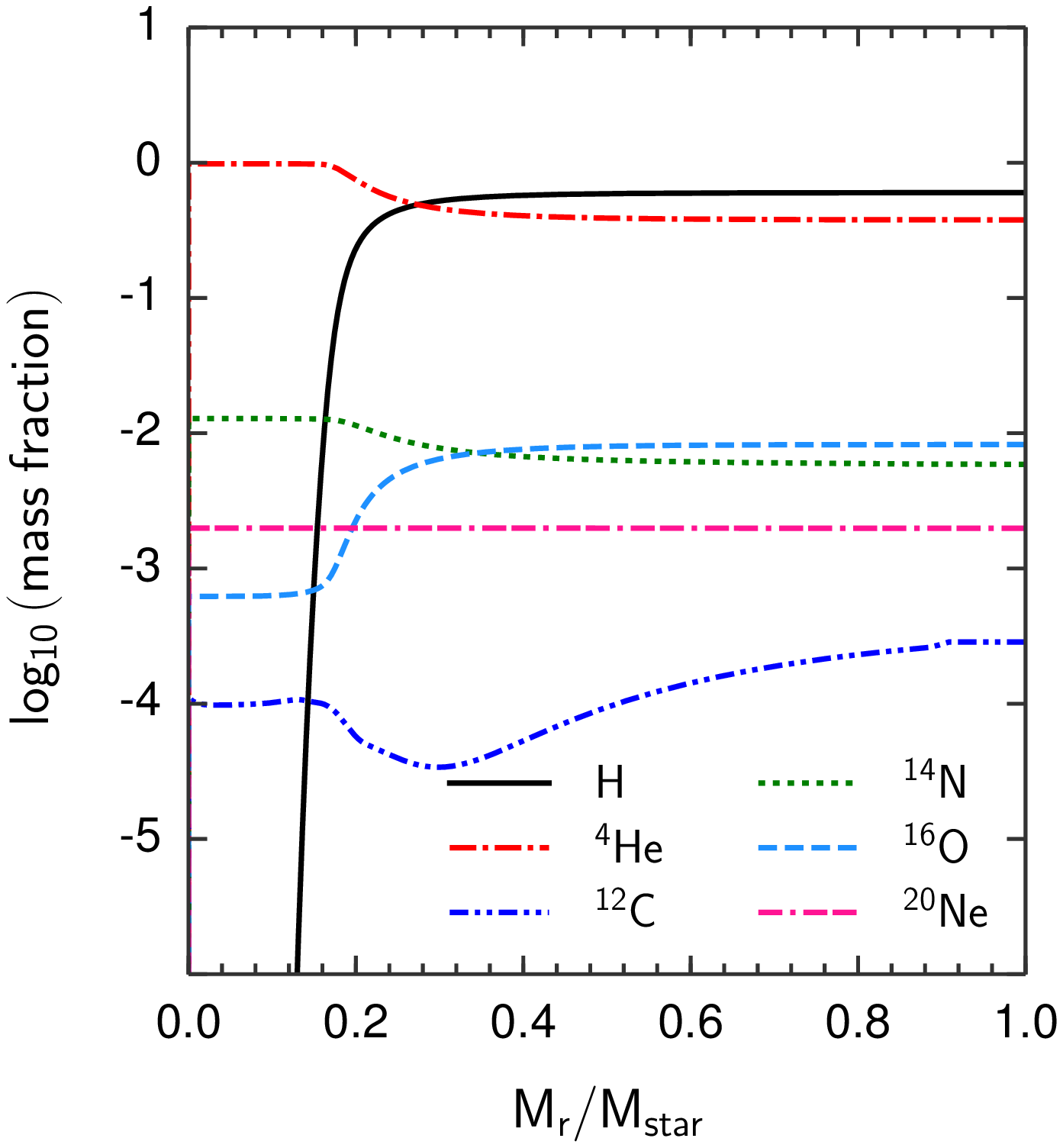}}
    \caption{Left panel: Hertzsprung-Russell diagram showing the evolutionary tracks of the primary (double dotted curve) and companion star (solid and single dotted curves) in the model ``M02'' (Table~\ref{Table:1}). Different evolutionary stages in Fig.~\ref{Fig:1} are marked with Roman numerals. A star symbol shows the moment of SN explosion. Right panel: chemical composition profiles of the HEMS donor star as a function of mass coordinate when the SN explodes.}
\label{Fig:2}
  \end{center}
\end{figure*}

\section{Numerical Methods}
\label{sec:code}

We use the Cambridge stellar evolution code {\sc STARS} \citep{Eggl71, Eggl72, Pols1995, Stancliffe2009} to trace the detailed binary evolution for the HEMS donor scenario. We start our calculation when a massive primary star evolves to a helium star. i.e., at stage $\mathrm{IV}$ in Fig.~\ref{Fig:1}, where the initial parameters of these binary systems are selected based on the results of \citetalias{Hach99} (see their Figure~1). Therefore, our initial binary systems are set to have a primary star of $1.0$--$1.8\,M_{\sun}$ and a secondary of $2.0$--$3.0\,M_{\sun}$ with an orbital period of $1.0$--$8.0\,\mathrm{d}$. Thermohaline mixing has been implemented into the code as described by \citet{Stancliffe2007}. The detailed structures of two components in the binary system are consistently solved in our calculation from stage $\mathrm{IV}$ to $\mathrm{VII}$. Once the helium primary star evolves to become a WD and the MS companion starts to fill its Roche lobe at stage $\mathrm{VII}$, instead of solving the detailed structure of the WD, we treat the WD as a point mass and follow the method of \citetalias{Hach99} to calculate the mass growth rate of the WD, $\dot{M}_{\rm{WD}}$. We set up the mass accumulation efficiency for hydrogen shell burning $\eta_{\rm{H}}$ as follows.
 \begin{equation}
    \label{eq:1}
\footnotesize
\begin{array}{l}
\eta_{\rm{H}} = \\[0.5em]
\left\{ \begin{array}{ll}
\dot{M}_{\rm{cr}}/|\dot{M}_{\rm{2}}|,  & (10^{-4}\,M_{\sun}\,\mathrm{yr^{-1}} > |\dot{M}_{\rm{2}}| > \dot{M}_{\rm{cr}})  \\[0.3em]
1, & (\dot{M}_{\rm{cr}} \geqslant \dot{M}_{\rm{2}} \geqslant 10^{-7}\,M_{\sun}\,\mathrm{yr^{-1}})\\[0.3em]
0, & (\dot{M}_{\rm{2}} < 1 \times10^{-7}\,M_{\sun}\,\mathrm{yr^{-1}})
\end{array} \right.
\end{array}
  \end{equation}
where $\dot{M}_{\mathrm{cr}} = 1.2\times10^{-6}(M_{\mathrm{WD}}/M_{\odot}-0.4)\,M_{\sun}\,\mathrm{yr^{-1}}$ is the critical accretion rate for stable hydrogen burning, $M_{\rm{WD}}$ is the mass of the accreting WD, $\dot{M}_{\rm{2}}$ is the mass transfer rate. The optically thick wind \citetalias{Hach99} is assumed to blow off all unprocessed material if $|\dot{M}_{\rm{2}}| > \dot{M}_{\rm{cr}}$, and the lost material is assumed to take away the specific orbital angular momentum of the accreting WD.

The mass accumulation efficiency for helium shell burning $\eta_{\rm{He}}$, is calculated based on \citet{Kato1999}: 
 \begin{equation}
    \label{eq:2}
\footnotesize
\begin{array}{l} 
\eta_{\mathrm{He}} = \\[0.5em]
\left\{\begin{array}{ll}
1,&(-5.9 \leqslant \mathrm{log}\,\dot{M}_{\mathrm{He}}\lesssim-5.0)\\[0.3em]
-0.175\,(\mathrm{log}\,\dot{M}_{\mathrm{He}}+5.35)^{2}+1.05,&(-7.8\leqslant\mathrm{log}\,\dot{M}_{\mathrm{He}}<-5.9)
\end{array}\right.
\end{array}
  \end{equation}

We assume the WD explodes as an SN Ia,  i.e., at stage $\mathrm{VIII}$ in Fig.~\ref{Fig:1}, when its  masses reaches the Chandrasekhar-mass limit (which we take as $1.4\,M_{\sun}$). Any rotation of the WD is not considered in our calculation here.

\section{Results}
\label{sec:results}

Figure~\ref{Fig:2} presents an example of our detailed binary evolution calculation for a binary system (i.e., model M02 in Table~\ref{Table:1}) consisting of a $1.2\,M_{\sun}$ primary star (a helium star) and a $2.0\,M_{\sun}$ secondary (a MS star). This binary system has an initial orbital period of $2.0\,\mathrm{days}$ at stage $\mathrm{IV}$. The helium star transfers its helium-rich material to the MS companion star as it undergoes the RLOF episode at stage V. As a result, about $0.31\,M_{\sun}$  of helium-rich material is accumulated onto the surface of the MS companion, and a binary system consisting of an 0.89$\,M_{\sun}$ WD and a helium-enriched MS companion (with a total mass of 2.31$\,M_{\sun}$) with an orbital period of about $3.2\,\mathrm{d}$ is formed (stage VI). The surface mass fraction of helium immediately after accretion is 0.98, but the accreted layer is rapidly mixed into the companion's interior via thermohaline convection. By the time equilibrium has been reached, the surface helium mass fraction is around 0.4. The companion star subsequently fills its Roche lobe, transferring material to the WD, which eventually reaches the Chandrasehkar mass, at which point we assume than an SN explosion occurs (stage VIII). Detailed composition profiles of different elements as a function of mass coordinate at the moment of SN explosion (stage VIII) are given in the right-hand panel of Fig.~\ref{Fig:2}. In general, mass transfer from the helium star results in between 0.1 and 0.7$\,M_{\sun}$ being transferred to the MS companion.

Rather than performing computationally expensive hydrodynamical simulations to determine, the amount of material stripped from the companion by the SN ejecta, we directly adopt the power-law relationships between total swept-up masses and the ratio of orbital separation to companion radius, $A_{\mathrm{orb}}/R_{2}$, given by past impact simulations for the MS companion star model to predict the amount of companion material that can be blown off by the SN explosion. Specifically for this work, Eq.~2 of \citealt{Liu12} and Eq.~4 of \citealt{Pan12} are used. For a given SN explosion model, it has been found that the ratio of $A_{\mathrm{orb}}/R_{2}$ is the most important factor to determine the total mass removed by the SN explosion (e.g., \citealt{Mari00, Pakm08, Liu12, Pan12}). Once the total removed mass is obtained, we further calculate the swept-up pure hydrogen mass based on the mass fraction profiles of hydrogen of the companion star at the moment of the explosion (Fig.~\ref{Fig:2}). Here, we assume that the outer layers of the star are symmetrically removed via interaction with the SN ejecta. We obtain that  the mass range of swept-up companion material is about $0.17$--$0.30\,M_{\sun}$ according to \citet{Liu12}, or about $0.2$--$0.4\,M_{\sun}$ according to \citet{Pan12}. Taking the mass fraction of hydrogen ($X_{\mathrm{H}}\approx0.55$) in the outer layers of the star into account, the amount of swept-up pure hydrogen in the HEMS donor scenario is about $0.10$--$0.17\,M_{\sun}$ according to \citet{Liu12}, or about $0.13$--$0.22\,M_{\sun}$ according to \citet{Pan12}. We list the predicted stripped hydrogen masses in Table~\ref{Table:1}.

\section{Discussion and Summary}
\label{sec:summary}

We find that the outer layers of the companion star in the HEMS donor scenario proposed by \citetalias{Hach99} are strongly helium-enriched when the SN explodes. However, our calculations still predict that the amount of swept-up hydrogen in the HEMS donor scenario is more massive than $0.1\,M_{\sun}$, which is higher than the observational constraints on the total hydrogen masses in SN Ia progenitor systems, which is expected to be $\lesssim0.001-0.06\,M_{\sun}$. If the HEMS donor channel is a significant route to SNe Ia formation (something that we shall determine from population synthesis calculations in forthcoming work), then the non-detection of hydrogen in these systems is a serious problem. 

To predict the amount of swept-up hydrogen, we directly adopt the power-law fits to recent hydrodynamical simulations for the MS companion star model \citep{Liu12, Pan12}. These models adopt the structure of a normal, hydrogen-rich MS star. The structure of the companion star in the HEMS donor scenario will be different.  It has been found that the details of companion structures can slightly affect the fitting parameters of these power-law relations (see Table~3 of \citealt{Liu12}). Therefore, we do not expect it will significantly affect the results of swept-up hydrogen masses expected in our calculations.

The observational upper limits on the amount of hydrogen in the progenitor systems are given based on the one-dimensional spectral modelling of \citet{Matt05} and \citet{Lund13}. As discussed in detail by \citet{Maguire2016}, uncertainties in the spectral modelling and observations will affect the stripped hydrogen mass predicted from the observations. For instance, the number of elements, ionization states and atomic levels included by the spectral synthesis models is limited. As discussed in \citet{Lund15}, this could lead to an underestimation of the line scattering and fluorescence. Also, whether the low-velocity hydrogen-rich material that is well confined within the radioactive material of SN explosion ejecta is sufficiently powered by radioactive heating to produce hydrogen emissions is still uncertain. In addition, the one-dimensional modelling of \citet{Matt05} and \citet{Lund13} was under an assumption of spherical symmetry and ignored any clumpiness in the SN ejecta. However, hydrodynamical models show that material is not swept-up from the companion star symmetrically. This asymmetric distribution of the swept-up companion material is expected to affect the shape and wavelength of the observed line profiles \citep{Maguire2016}. If clumping occurs, it might lead to the swept-up hydrogen present in the ejecta not being observable, and thus we would underestimate the hydrogen mass. Therefore, multidimensional radiative transfer calculations, including the structure resulting from the interaction of the SN ejecta with the companion star are needed.

The results in Table~\ref{Table:1} give the total swept-up hydrogen masses due to both the ablation (heating) and stripping (momentum transfer) though the interaction of the SN ejecta with a companion star. Generally, the ablated material moves relatively slowly, which means it will be well confined within the radioactive material, have fairly high-density and will result in a narrow line profile that would be  more likely to be detected. However, the stripped material has the opposite features, and would be much less likely to be detectable \citep{Lund15}. Hydrodynamical simulations show that ablation removes $\gtrsim50\%$ of the material from the companion star. Therefore, the ablated hydrogen masses in the HEMS donor scenario is about $\gtrsim0.05$--$0.085\,M_{\sun}$ according to \citet{Liu12} or $\gtrsim0.065$--$0.11\,M_{\sun}$ according to \citet{Pan12}. The former prediction has some overlap with the observational mass limits, and given the uncertainties in the observational prediction, there is hope that these values could ultimately be reconciled. Finally, the observational limits are quite sensitive to the velocity of the stripped or ablated gas \citep{Lund15}, a tighter constraint on the hydrogen masses in SNe Ia can only be placed by performing more detailed spectral modelling as well as looking at a significantly larger sample of nebular spectra of SNe Ia.

In summary, current observations suggest that the amount of hydrogen in the progenitor systems of SNe Ia is $\lesssim0.001$-$0.06\,M_{\sun}$. However, a more massive hydrogen mass of $\gtrsim0.1\,M_{\sun}$ than the observational limits is expected to be blown off by the SN explosion in the HEMS donor scenario based on our self-consistent binary evolution calculations, which cannot provide a way to explain the lack of hydrogen lines in late-time spectra of SNe Ia. This presents a serious challenge to the HEMS donor scenario for producing SNe Ia.

\section*{Acknowledgements}
We thank the anonymous referee for his/her valuable comments that helped to improve the paper. ZWL thanks Peter Lundqvist for his helpful discussions about the spectral synthesis modelling. This work is supported by the Alexander von Humboldt Foundation. R.J.S. is the recipient of a Sofja Kovalevskaja Award from the Alexander von Humboldt Foundation.

\bibliographystyle{mnras}

\bibliography{ref}

\begin{thebibliography}{}
\makeatletter
\relax
\def\mn@urlcharsother{\let\do\@makeother \do\$\do\&\do\#\do\^\do\_\do\%\do\~}
\def\mn@doi{\begingroup\mn@urlcharsother \@ifnextchar [ {\mn@doi@}
  {\mn@doi@[]}}
\def\mn@doi@[#1]#2{\def\@tempa{#1}\ifx\@tempa\@empty \href
  {http://dx.doi.org/#2} {doi:#2}\else \href {http://dx.doi.org/#2} {#1}\fi
  \endgroup}
\def\mn@eprint#1#2{\mn@eprint@#1:#2::\@nil}
\def\mn@eprint@arXiv#1{\href {http://arxiv.org/abs/#1} {{\tt arXiv:#1}}}
\def\mn@eprint@dblp#1{\href {http://dblp.uni-trier.de/rec/bibtex/#1.xml}
  {dblp:#1}}
\def\mn@eprint@#1:#2:#3:#4\@nil{\def\@tempa {#1}\def\@tempb {#2}\def\@tempc
  {#3}\ifx \@tempc \@empty \let \@tempc \@tempb \let \@tempb \@tempa \fi \ifx
  \@tempb \@empty \def\@tempb {arXiv}\fi \@ifundefined
  {mn@eprint@\@tempb}{\@tempb:\@tempc}{\expandafter \expandafter \csname
  mn@eprint@\@tempb\endcsname \expandafter{\@tempc}}}

\bibitem[\protect\citeauthoryear{{Aldering} et~al.,}{{Aldering}
  et~al.}{2006}]{Aldering2006}
{Aldering} G.,  et~al., 2006, \mn@doi [\apj] {10.1086/507020}, \href
  {http://adsabs.harvard.edu/abs/2006ApJ...650..510A} {650, 510}

\bibitem[\protect\citeauthoryear{{Boehner}, {Plewa}  \& {Langer}}{{Boehner}
  et~al.}{2017}]{Boehner2017}
{Boehner} P.,  {Plewa} T.,   {Langer} N.,  2017, \mn@doi [\mnras]
  {10.1093/mnras/stw2737}, \href
  {http://adsabs.harvard.edu/abs/2017MNRAS.465.2060B} {465, 2060}

\bibitem[\protect\citeauthoryear{{Dilday} et~al.,}{{Dilday}
  et~al.}{2012}]{Dilday2012}
{Dilday} B.,  et~al., 2012, \mn@doi [Science] {10.1126/science.1219164}, \href
  {http://adsabs.harvard.edu/abs/2012Sci...337..942D} {337, 942}

\bibitem[\protect\citeauthoryear{{Eggleton}}{{Eggleton}}{1971}]{Eggl71}
{Eggleton} P.~P.,  1971, \mnras, \href
  {http://adsabs.harvard.edu/abs/1971MNRAS.151..351E} {151, 351}

\bibitem[\protect\citeauthoryear{{Eggleton}}{{Eggleton}}{1972}]{Eggl72}
{Eggleton} P.~P.,  1972, \mnras, \href
  {http://adsabs.harvard.edu/abs/1972MNRAS.156..361E} {156, 361}

\bibitem[\protect\citeauthoryear{{Graham}, {Nugent}, {Sullivan}, {Filippenko},
  {Cenko}, {Silverman}, {Clubb}  \& {Zheng}}{{Graham}
  et~al.}{2015}]{Graham2015}
{Graham} M.~L.,  {Nugent} P.~E.,  {Sullivan} M.,  {Filippenko} A.~V.,  {Cenko}
  S.~B.,  {Silverman} J.~M.,  {Clubb} K.~I.,   {Zheng} W.,  2015, \mn@doi
  [\mnras] {10.1093/mnras/stv1888}, \href
  {http://adsabs.harvard.edu/abs/2015MNRAS.454.1948G} {454, 1948}

\bibitem[\protect\citeauthoryear{{Hachisu}, {Kato}, {Nomoto}  \&
  {Umeda}}{{Hachisu} et~al.}{1999}]{Hach99}
{Hachisu} I.,  {Kato} M.,  {Nomoto} K.,   {Umeda} H.,  1999, \mn@doi [\apj]
  {10.1086/307370}, \href {http://adsabs.harvard.edu/abs/1999ApJ...519..314H}
  {519, 314}

\bibitem[\protect\citeauthoryear{{Hamuy} et~al.,}{{Hamuy}
  et~al.}{2003}]{Hamuy2003}
{Hamuy} M.,  et~al., 2003, \mn@doi [\nat] {10.1038/nature01854}, \href
  {http://adsabs.harvard.edu/abs/2003Natur.424..651H} {424, 651}

\bibitem[\protect\citeauthoryear{{Han} \& {Podsiadlowski}}{{Han} \&
  {Podsiadlowski}}{2004}]{Han04}
{Han} Z.,  {Podsiadlowski} P.,  2004, \mn@doi [\mnras]
  {10.1111/j.1365-2966.2004.07713.x}, \href
  {http://adsabs.harvard.edu/abs/2004MNRAS.350.1301H} {350, 1301}

\bibitem[\protect\citeauthoryear{{Hillebrandt} \& {Niemeyer}}{{Hillebrandt} \&
  {Niemeyer}}{2000}]{Hillebrandt2000}
{Hillebrandt} W.,  {Niemeyer} J.~C.,  2000, \mn@doi [\araa]
  {10.1146/annurev.astro.38.1.191}, \href
  {http://adsabs.harvard.edu/abs/2000ARA%26A..38..191H} {38, 191}

\bibitem[\protect\citeauthoryear{{Iben} \& {Tutukov}}{{Iben} \&
  {Tutukov}}{1984}]{Iben84}
{Iben} Jr. I.,  {Tutukov} A.~V.,  1984, \mn@doi [\apj] {10.1086/162455}, \href
  {http://adsabs.harvard.edu/abs/1984ApJ...284..719I} {284, 719}

\bibitem[\protect\citeauthoryear{{Kato} \& {Hachisu}}{{Kato} \&
  {Hachisu}}{1999}]{Kato1999}
{Kato} M.,  {Hachisu} I.,  1999, \mn@doi [\apjl] {10.1086/311893}, \href
  {http://adsabs.harvard.edu/abs/1999ApJ...513L..41K} {513, L41}

\bibitem[\protect\citeauthoryear{{Kippenhahn}, {Ruschenplatt}  \&
  {Thomas}}{{Kippenhahn} et~al.}{1980}]{Kippenhahn1980}
{Kippenhahn} R.,  {Ruschenplatt} G.,   {Thomas} H.-C.,  1980, \aap, \href
  {http://adsabs.harvard.edu/abs/1980A%26A....91..175K} {91, 175}

\bibitem[\protect\citeauthoryear{{Leonard}}{{Leonard}}{2007}]{Leon07}
{Leonard} D.~C.,  2007, \mn@doi [\apj] {10.1086/522367}, \href
  {http://adsabs.harvard.edu/abs/2007ApJ...670.1275L} {670, 1275}

\bibitem[\protect\citeauthoryear{{Liu}, {Pakmor}, {R{\"o}pke}, {Edelmann},
  {Wang}, {Kromer}, {Hillebrandt}  \& {Han}}{{Liu} et~al.}{2012}]{Liu12}
{Liu} Z.~W.,  {Pakmor} R.,  {R{\"o}pke} F.~K.,  {Edelmann} P.,  {Wang} B.,
  {Kromer} M.,  {Hillebrandt} W.,   {Han} Z.~W.,  2012, \mn@doi [\aap]
  {10.1051/0004-6361/201219357}, \href
  {http://adsabs.harvard.edu/abs/2012A%26A...548A...2L} {548, A2}

\bibitem[\protect\citeauthoryear{{Liu}, {Pakmor}, {R{\"o}pke}, {Edelmann},
  {Hillebrandt}, {Kerzendorf}, {Wang}  \& {Han}}{{Liu} et~al.}{2013a}]{Liu13a}
{Liu} Z.-W.,  {Pakmor} R.,  {R{\"o}pke} F.~K.,  {Edelmann} P.,  {Hillebrandt}
  W.,  {Kerzendorf} W.~E.,  {Wang} B.,   {Han} Z.~W.,  2013a, \mn@doi [\aap]
  {10.1051/0004-6361/201220903}, \href
  {http://adsabs.harvard.edu/abs/2013A%26A...554A.109L} {554, A109}

\bibitem[\protect\citeauthoryear{{Liu}, {Kromer}, {Fink}, {Pakmor},
  {R{\"o}pke}, {Chen}, {Wang}  \& {Han}}{{Liu} et~al.}{2013b}]{Liu13b}
{Liu} Z.-W.,  {Kromer} M.,  {Fink} M.,  {Pakmor} R.,  {R{\"o}pke} F.~K.,
  {Chen} X.~F.,  {Wang} B.,   {Han} Z.~W.,  2013b, \mn@doi [\apj]
  {10.1088/0004-637X/778/2/121}, \href
  {http://adsabs.harvard.edu/abs/2013ApJ...778..121L} {778, 121}

\bibitem[\protect\citeauthoryear{{Lundqvist} et~al.,}{{Lundqvist}
  et~al.}{2013}]{Lund13}
{Lundqvist} P.,  et~al., 2013, \mn@doi [\mnras] {10.1093/mnras/stt1303}, \href
  {http://adsabs.harvard.edu/abs/2013MNRAS.435..329L} {435, 329}

\bibitem[\protect\citeauthoryear{{Lundqvist} et~al.,}{{Lundqvist}
  et~al.}{2015}]{Lund15}
{Lundqvist} P.,  et~al., 2015, \mn@doi [\aap] {10.1051/0004-6361/201525719},
  \href {http://adsabs.harvard.edu/abs/2015A%26A...577A..39L} {577, A39}

\bibitem[\protect\citeauthoryear{{Maguire}, {Taubenberger}, {Sullivan}  \&
  {Mazzali}}{{Maguire} et~al.}{2016}]{Maguire2016}
{Maguire} K.,  {Taubenberger} S.,  {Sullivan} M.,   {Mazzali} P.~A.,  2016,
  \mn@doi [\mnras] {10.1093/mnras/stv2991}, \href
  {http://adsabs.harvard.edu/abs/2016MNRAS.457.3254M} {457, 3254}

\bibitem[\protect\citeauthoryear{{Maoz}, {Mannucci}  \& {Nelemans}}{{Maoz}
  et~al.}{2014}]{Maoz13}
{Maoz} D.,  {Mannucci} F.,   {Nelemans} G.,  2014, \mn@doi [\araa]
  {10.1146/annurev-astro-082812-141031}, \href
  {http://adsabs.harvard.edu/abs/2014ARA%26A..52..107M} {52, 107}

\bibitem[\protect\citeauthoryear{{Marietta}, {Burrows}  \&
  {Fryxell}}{{Marietta} et~al.}{2000}]{Mari00}
{Marietta} E.,  {Burrows} A.,   {Fryxell} B.,  2000, \mn@doi [\apjs]
  {10.1086/313392}, \href {http://adsabs.harvard.edu/abs/2000ApJS..128..615M}
  {128, 615}

\bibitem[\protect\citeauthoryear{{Mattila}, {Lundqvist}, {Sollerman}, {Kozma},
  {Baron}, {Fransson}, {Leibundgut}  \& {Nomoto}}{{Mattila}
  et~al.}{2005}]{Matt05}
{Mattila} S.,  {Lundqvist} P.,  {Sollerman} J.,  {Kozma} C.,  {Baron} E.,
  {Fransson} C.,  {Leibundgut} B.,   {Nomoto} K.,  2005, \mn@doi [\aap]
  {10.1051/0004-6361:20052731}, \href
  {http://adsabs.harvard.edu/abs/2005A%26A...443..649M} {443, 649}

\bibitem[\protect\citeauthoryear{{Pakmor}, {R{\"o}pke}, {Weiss}  \&
  {Hillebrandt}}{{Pakmor} et~al.}{2008}]{Pakm08}
{Pakmor} R.,  {R{\"o}pke} F.~K.,  {Weiss} A.,   {Hillebrandt} W.,  2008,
  \mn@doi [\aap] {10.1051/0004-6361:200810456}, \href
  {http://adsabs.harvard.edu/abs/2008A%26A...489..943P} {489, 943}

\bibitem[\protect\citeauthoryear{{Pan}, {Ricker}  \& {Taam}}{{Pan}
  et~al.}{2012}]{Pan12}
{Pan} K.-C.,  {Ricker} P.~M.,   {Taam} R.~E.,  2012, \mn@doi [\apj]
  {10.1088/0004-637X/750/2/151}, \href
  {http://adsabs.harvard.edu/abs/2012ApJ...750..151P} {750, 151}

\bibitem[\protect\citeauthoryear{{Pols}, {Tout}, {Eggleton}  \& {Han}}{{Pols}
  et~al.}{1995}]{Pols1995}
{Pols} O.~R.,  {Tout} C.~A.,  {Eggleton} P.~P.,   {Han} Z.,  1995, \mn@doi
  [\mnras] {10.1093/mnras/274.3.964}, \href
  {http://adsabs.harvard.edu/abs/1995MNRAS.274..964P} {274, 964}

\bibitem[\protect\citeauthoryear{{Shappee}, {Stanek}, {Pogge}  \&
  {Garnavich}}{{Shappee} et~al.}{2013}]{Shap13}
{Shappee} B.~J.,  {Stanek} K.~Z.,  {Pogge} R.~W.,   {Garnavich} P.~M.,  2013,
  \mn@doi [\apjl] {10.1088/2041-8205/762/1/L5}, \href
  {http://adsabs.harvard.edu/abs/2013ApJ...762L...5S} {762, L5}

\bibitem[\protect\citeauthoryear{{Shappee}, {Piro}, {Stanek}, {Patel},
  {Margutti}, {Lipunov}  \& {Pogge}}{{Shappee} et~al.}{2016}]{shappee16}
{Shappee} B.~J.,  {Piro} A.~L.,  {Stanek} K.~Z.,  {Patel} S.~G.,  {Margutti}
  R.~A.,  {Lipunov} V.~M.,   {Pogge} R.~W.,  2016, ArXiv:1610.07601, \href
  {http://adsabs.harvard.edu/abs/2016arXiv161007601S} {}

\bibitem[\protect\citeauthoryear{{Siess}}{{Siess}}{2009}]{Siess2009}
{Siess} L.,  2009, \mn@doi [\aap] {10.1051/0004-6361/200811362}, \href
  {http://adsabs.harvard.edu/abs/2009A%26A...497..463S} {497, 463}

\bibitem[\protect\citeauthoryear{{Silverman} et~al.,}{{Silverman}
  et~al.}{2013}]{Silverman2013}
{Silverman} J.~M.,  et~al., 2013, \mn@doi [\apjs] {10.1088/0067-0049/207/1/3},
  \href {http://adsabs.harvard.edu/abs/2013ApJS..207....3S} {207, 3}

\bibitem[\protect\citeauthoryear{{Stancliffe}}{{Stancliffe}}{2010}]{Stancliffe2010}
{Stancliffe} R.~J.,  2010, \mn@doi [\mnras] {10.1111/j.1365-2966.2009.16150.x},
  \href {http://adsabs.harvard.edu/abs/2010MNRAS.403..505S} {403, 505}

\bibitem[\protect\citeauthoryear{{Stancliffe} \& {Eldridge}}{{Stancliffe} \&
  {Eldridge}}{2009}]{Stancliffe2009}
{Stancliffe} R.~J.,  {Eldridge} J.~J.,  2009, \mn@doi [\mnras]
  {10.1111/j.1365-2966.2009.14849.x}, \href
  {http://adsabs.harvard.edu/abs/2009MNRAS.396.1699S} {396, 1699}

\bibitem[\protect\citeauthoryear{{Stancliffe} \& {Glebbeek}}{{Stancliffe} \&
  {Glebbeek}}{2008}]{Stancliffe2008}
{Stancliffe} R.~J.,  {Glebbeek} E.,  2008, \mn@doi [\mnras]
  {10.1111/j.1365-2966.2008.13700.x}, \href
  {http://adsabs.harvard.edu/abs/2008MNRAS.389.1828S} {389, 1828}

\bibitem[\protect\citeauthoryear{{Stancliffe}, {Glebbeek}, {Izzard}  \&
  {Pols}}{{Stancliffe} et~al.}{2007}]{Stancliffe2007}
{Stancliffe} R.~J.,  {Glebbeek} E.,  {Izzard} R.~G.,   {Pols} O.~R.,  2007,
  \mn@doi [\aap] {10.1051/0004-6361:20066891}, \href
  {http://adsabs.harvard.edu/abs/2007A%26A...464L..57S} {464, L57}

\bibitem[\protect\citeauthoryear{{Webbink}}{{Webbink}}{1984}]{Webb84}
{Webbink} R.~F.,  1984, \mn@doi [\apj] {10.1086/161701}, \href
  {http://adsabs.harvard.edu/abs/1984ApJ...277..355W} {277, 355}

\bibitem[\protect\citeauthoryear{{Wellstein}, {Langer}  \& {Braun}}{{Wellstein}
  et~al.}{2001}]{Wellstein2001}
{Wellstein} S.,  {Langer} N.,   {Braun} H.,  2001, \mn@doi [\aap]
  {10.1051/0004-6361:20010151}, \href
  {http://adsabs.harvard.edu/abs/2001A%26A...369..939W} {369, 939}

\bibitem[\protect\citeauthoryear{{Wheeler}, {Lecar}  \& {McKee}}{{Wheeler}
  et~al.}{1975}]{Whee75}
{Wheeler} J.~C.,  {Lecar} M.,   {McKee} C.~F.,  1975, \mn@doi [\apj]
  {10.1086/153771}, \href {http://adsabs.harvard.edu/abs/1975ApJ...200..145W}
  {200, 145}

\bibitem[\protect\citeauthoryear{{Whelan} \& {Iben}}{{Whelan} \&
  {Iben}}{1973}]{Whel73}
{Whelan} J.,  {Iben} Jr. I.,  1973, \mn@doi [\apj] {10.1086/152565}, \href
  {http://adsabs.harvard.edu/abs/1973ApJ...186.1007W} {186, 1007}

\bibitem[\protect\citeauthoryear{{Wood-Vasey}, {Wang}  \&
  {Aldering}}{{Wood-Vasey} et~al.}{2004}]{Wood-Vasey2004}
{Wood-Vasey} W.~M.,  {Wang} L.,   {Aldering} G.,  2004, \mn@doi [\apj]
  {10.1086/424826}, \href {http://adsabs.harvard.edu/abs/2004ApJ...616..339W}
  {616, 339}

\makeatother
\end{thebibliography}

\end{document}